*Materialising contexts: virtual soundscapes for real-world exploration*

# Laurence Cliffe, James Mansell, Chris Greenhalgh & Adrian Hazzard

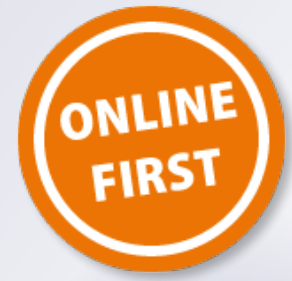

**ORIGINAL PAPER**

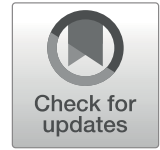

# Materialising contexts: virtual soundscapes for real-world exploration

Laurence Cliffe[1] · James Mansell[1] · Chris Greenhalgh[1] · Adrian Hazzard[1]

**Abstract**
This article presents the results of a study based on a group of participants' interactions with an experimental sound installation at the National Science and Media Museum in Bradford, UK. The installation used audio augmented reality to attach virtual sound sources to a vintage radio receiver from the museum's collection, with a view to understanding the potentials of this technology for promoting exploration and engagement within museums and galleries. We employ a practice-based design ethnography, including a thematic analysis of our participants' interactions with spatialised interactive audio, and present an identified sequence of interactional phases. We discuss how audio augmented artefacts can communicate and engage visitors beyond their traditional confines of line-of-sight, and how visitors can be drawn to engage further, beyond the realm of their original encounter. Finally, we provide evidence of how contextualised and embodied interactions, along with authentic audio reproduction, evoked personal memories associated with our museum artefact, and how this can promote interest in the acquisition of declarative knowledge. Additionally, through the adoption of a functional and theoretical aura-based model, we present ways in which this could be achieved, and, overall, we demonstrate a material object's potential role as an interface for engaging users with, and contextualising, immaterial digital audio archival content.

## 1 Introduction

The audio augmented reality (AAR) installation presented here, along with the subsequent study of its deployment, aims to investigate and access the potentials and challenges involved in utilising such technology as a means of promoting visitor exploration and engagement with physical museum and gallery-based artefacts, along with related digital audio archive material. Initially, we provide some theoretical background against which the project has been developed, then outline some additional background examples of culturally applied AAR projects in the form of related work.

✉ Laurence Cliffe
laurence.cliffe@nottingham.ac.uk

James Mansell
james.mansell@nottingham.ac.uk

Chris Greenhalgh
christopher.greenhalgh@nottingham.ac.uk

Adrian Hazzard
adrian.hazzard@nottingham.ac.uk

[1] University of Nottingham, Nottingham, UK

After a description of our practice-based research through design approach, a detailed technical description of the AAR installation is provided, including its authorship, development and application. Our study is described, our findings are presented and a discussion is included based around the system's perceived ability to extend the communicative potential of the museum and gallery object, and in relation to affording primacy to the sonic, rather than the visual. Finally, we present our conclusions and outline some potential future avenues for exploration.

Within the context of this article, AAR is considered a virtual audio augmentation of the physical and visual reality, or the physical artefact. In an approach similar to [2, 29, 33], a virtual audio soundscape currently replaces the ambient acoustic reality of the location, rather than mixing with it, a mixed reality experience is therefore realised through the meeting of physical artefact and virtual audio.

## 2 Background

In the academic literature on museums, sound has been identified as having the potential to give exhibitions emotional power [7] and to generate a multiplicity of interpretative

Springer

perspective [5, 19]. The argument, in short, is that sonic exhibitions might help us to break from the truth effects of visual and textual storytelling and all of the asymmetrical power relations that they have been said to produce. Such ideas are especially evident within Foucauldian critiques of museums, and sonic exhibitions may help open the ground for visitors to 'poach' what they need from exhibitions, to borrow Boon's paraphrasing [5] of Michel de Certeau. Museums have enthusiastically embraced the challenge of sound, identifying its potential to produce more entertaining exhibitions, most notably in order to deal with auditory subject matter as in the case of the V&A's exhibitions 'David Bowie Is' and 'Pink Floyd: Their Mortal Remains' both of which provided a fully soundtracked experience on headphones. Equally of note is the Wellcome Collection's less obviously crowd-pleasing 2016 exhibition 'This is a Voice' which used installed sound, mainly via contemporary art commissions, to tell the scientific, medical and cultural story of the human voice. This trajectory has established sound as an interpretation tactic in museums. However, there remains a live question about how to approach sound itself as an object of display.

There is also the challenge posed by the curious practice of collecting media technology and media content separately. The national sound archive is now held at the British Library, isolated from the objects which once created and replayed recorded sound held largely at Science Museum and its regional branches, especially the National Science and Media Museum in Bradford. In response to the rapidly deteriorating physical state of British Library sound archive materials and others like it in regional collections, the Library has embarked on an ambitious programme of digitisation known as 'Unlocking Our Sound Heritage' [6], though there remains little sense of what public use will be made of this digital archive once it is made available. From a silenced collection of sound technology hardware to an abundant, even noisy, digital sound archive, there is at present little strategy or consensus about what might be termed 'sonic engagement'—the practice of engaging the public in the history of hearing, listening and sound. The question of what sonic engagement should mean and how it should be achieved in the context of museums of science and technology was taken up by the Gallery Listening Sessions project at the National Science and Media Museum.

## 3 Related work

In addition to the exhibition-based audio experiences outlined within the introduction of this article, there are a number of other related projects that provide useful reference points, particularly in relation to a similar applied use of AAR.

Zimmerman and Lorenz's LISTEN system [33] provides an excellent example of the capabilities of AAR within the context of a cultural institution. The LISTEN project, which they describe as 'an attempt to make use of the inherent everyday integration of aural and visual perception', delivers a personalised and interactive location-based audio experience based on an adaptive system model. It does this by tracking aspects of the visitors behaviour (which artworks have been visited, how long were they visited for etc.) to assign the visitor a behavioural model and adjust the delivery of audio content accordingly. The LISTEN system relies on a substantial technical background infrastructure to realise this personalised and invisible technical front-end experience for the visitor, who can wander freely through the exhibition space with just a set of customised headphones. LISTEN also introduces the concept of the attractor sound, which, based on the visitor's personalised profile model, suggests other nearby artworks to the visitor that may be of interest to them via spatially located audio prompts. Furthermore, LISTEN characterises many of the key differences between the usual audio guide experience and an interactive, adaptive and immersive approach. These include binaurally rendered, three-dimensional surround sound based on the listener's movement and the delivery of related audio content based on the listener's proximity to an exhibit. The authors report that two-thirds of participants rated their experience with the LISTEN system as being 'enriching', and clear positive feedback was gathered in relation to the combination of artwork and auditory information realised through the system.

Hazzard et al's 'The Rough Mile' [17], where pre-recorded audio is used to augment a specific outdoor location, and Sikora et al's archaeological AAR experience [29], where pre-recorded audio is used to augment locations in and around an archaeological site, could both be categorised as examples of transformative soundscapes. In both of these examples, audio sources are used to reframe, rather than to directly compliment, the context of the locative experience. In the case of Sikora et al's AAR experience, this change of context is from rural to urban; in 'The Rough Mile', this change of context is from city centre to fictional narrative. Being outdoor experiences, both rely on GPS technology for determining the position of the user within the physical landscape. In Sikora et al's AAR experience, the listener's GPS coordinate values are plotted on a virtually authored representation of the landscape based on satellite imagery, onto which are placed virtual sound sources for the user to encounter in the real world. A similar authoring approach is taken by the system presented here; though being for an indoor experience, it relies on a custom indoor positioning approach rather than GPS for determining the position of the user within virtual and physical space.

Seidenari et al's work on an automatic context-aware audio museum guide [28] demonstrates how a combination of both context modelling and artwork detection work together to influence the playback of audio descriptions. It also shows how the current object of the visitor's focus is determined by a wearable camera-based object recognition system. Additionally, the inclusion of speech detection within Seidenari et al's context-aware audio guide suggests a desire for users of such systems to maintain the ability to socially interact with their co-visitors, or rather it tries to ensure that visitors can still talk to other visitors. This ability is maintained in addition to an understanding that personalisation is a key factor in enabling museums to talk with visitors, rather than talking to them. Seidenari et al. report an above average feedback score for their experience design [28] which, in the main part, is attributed to the increased user agency that their *SeeForMe* system affords over more traditional audio guides. This evaluation is also attributed to the fluidity of the experience, and the ability of their system to make users aware of other artworks around them.

The project presented here, where an AAR installation environment has been created using a vintage radio receiver and contemporaneous archive radio broadcast recordings, has subsequently been found to be similar to one referenced by Bijsterveld [4] and presented in detail by Mortensen and Vestergaard [23] within what they term a listening exhibition curated at the Media Museum in Odense, Demark, in 2012 titled 'You are what you hear'. Through the implementation of their Exaudimus system [23], Mortensen and Vestergaard propose a way of exhibiting and interfacing with radio heritage which has been enabled by the digitisation of analogue audio archive content by the Danish Broadcasting Corporation. Within this approach, we see how, through authorship and embodied visitor interaction, the exhibition demonstrates potential as an accessible and immersive interface to the sound archive itself.

We can imagine the audible output of the two projects to be of a similar nature, given the similar context and type of physical and virtual audio artefacts used. But the application of different technological solutions within each AAR system and the apparent absence of three-dimensional audio spatialisation within the Exaudimus system, along with differences in the material contextualisation of the audio content (listening situation verses direct augmentation of the sound artefact), denote the issues around both authorship and user experience being very much different.

Whereas the Exaudimus system [23] utilises a multiple fixed camera tracking system, which tracks different coloured lights mounted on top of the user's headphones in order to determine the position of a specific user within the installation environment, the prototype presented here employs a single handheld mobile camera-based tracking system along with Simultaneous Localisation and Mapping (SLAM) [32] to determine the user's physical location in space in relation to the virtual audio sources.

Although Mortensen and Vestergaard report some success in generating engagement with the audio archive content contained within the exhibition, significant issues arose around initiating interaction with, and triggering the playback of the audio archive content. The authors attribute this to what they term Cultural Constraints, the reluctance of visitors to touch, pick up and directly interact with physical objects within a gallery environment, something which goes against normal behaviour within such a context. Unfortunately, the triggering of archive audio playback was largely dependent on such direct interactions with the constructed listening situations within the exhibition.

## 4 Approach and methodology

The project employed a practice-based research through design approach where a series of iterative prototype interactive sound installations were realised through a cyclical process of development, deployment, study, analysis and redevelopment [3]. Both experts and prospective audiences were invited to participate and interact with the installation, and these interactions were observed, recorded and thematically analysed in accordance with recognised ethnomethodological techniques, including the development of thick descriptions and a detailed understanding of the machinery of interaction [3, 12]. Additional data in relation to participant experiences were obtained from a post-participatory questionnaire and informal discussions.

This approach to developing novel interactive experiences and then applying design ethnography as a methodology for its study within the context for which it has been designed can be closely associated with what Benford et al. [3] term as 'Performance-led research in the wild'. Furthermore, Benford et al. suggest that many of the techniques that artists adopt within the creation and deployment of interactive artworks can prove useful within the development of cultural experiences in general. Within such an approach, what is of ultimate concern is the generation of theory as a product of the creative process and application of the research-related project. This approach is also recognised by Gaver and Bowers [13] as a potential route to the discovery of generalizable theory within a field. In relation to the adoption of such an approach, it is worth noting that both Benford et al. and Gaver and Bowers outline the importance of maintaining artistic integrity within such a project and warn of the dangers of the forced application of theory on the creative process. This point is made in relation to the fear of losing that which makes a practice-based approach an important contributing element to the research.

## 5 System description

The current prototype installation is delivered to listeners though a set of stereo, closed-cup, over-ear headphones connected to a smartphone. Installed on the smartphone is an application that is authored using the Unity Game Engine [31], FMOD Studio adaptive game audio authoring tool [11] and the augmented reality library Vuforia [26] (see Fig. 1).

Each sound source either has an audio logic script attached to it or is attached to an FMOD event, which is provided with the current distance and orientation values of the listener in relation to it, which it uses to control the delivery of the audio source to the listener. This includes its spatial position within the virtual soundscape, based on the listener's orientation in relation to the virtual sound source and the real-world object, and its attenuation within the virtual soundscape, based on the listener's distance from the virtual sound source and the real-world object. The spatial position and the attenuation of the sound source within the stereo binaural mix of the virtual soundscape are the primary audio logic parameters which all the sound sources contain in order to place them within, and construct, a convincing and viable interactive and virtual three-dimensional soundscape. Based on these orientation and distance values, other audio logic events can be scripted, such as the delivery of different audio files, or sections of an audio file, based on the listener's position in relation to the source.

The Resonance Audio [14] VR Spatialiser plugin was used within FMOD and selected as the preferred audio spatialiser within the Unity project. An exponential attenuation curve was used with a maximum distance attenuation range of two meters (with the exception of the background static which had a range of three meters) along with a reference volume of 0.2 m for all the audio sources.

The Vuforia SDK [26] was adopted as a means to realise an image recognition and tracking feature within the system that was useable from both an authoring and curatorial perspective in a variety of locations. This decision was informed and inspired by the artwork detection project presented by Seidenari et al. [9]. Along with artwork recognition, the use of image recognition and tracking technology presented opportunities for the development of an Indoor Positioning System (IPS). The Vuforia SDK enables the development of mobile augmented reality applications that use computer vision technology to recognise and track image targets and three-dimensional objects in real-time, and is compatible with both the iOS and Android mobile application platforms. The Vuforia Engine's camera-based object recognition and tracking capabilities not only facilitate the recognition of the artwork and artefacts to which virtual audio sources can be associated but also additionally enable the implementation of an IPS where the mobile listener's angle and distance can be determined in relation to tracked, stationary two or three-dimensional objects.

Through an authoring approach similar to the one presented in the LISTEN system by Zimmerman and Lorenz [33], where a world model is combined with a locative model, we can determine our listener's position both in the physical and virtual environment of the experience. Within Zimmerman and Lorenz's LISTEN system, the world model contains geometric information relating to the physical real-world environment and the objects within it, which it describes as the visitor moves and interacts with the system. On the other hand, the location model defines areas of interaction within the world model and enables the system to determine the visitor's location and head orientation by mapping their position to predetermined virtual zones within the space and their position in relation to object identifiers.

Within the prototype system presented here, our locative model is authored within Unity as zones of space of a specified size and shape, situated at specific coordinates in three-dimensional space in relation to a unique and recognisable image target. The Vuforia SDK acts as our world model, which it creates on-the-fly, recognising and tracking the location of the image target in the physical environment. Because our listener is holding the camera, the system knows the listeners' position and bodily orientation in relation to the tracked image and therefore can determine the listeners' position and orientation in relation to our authored zones of space.

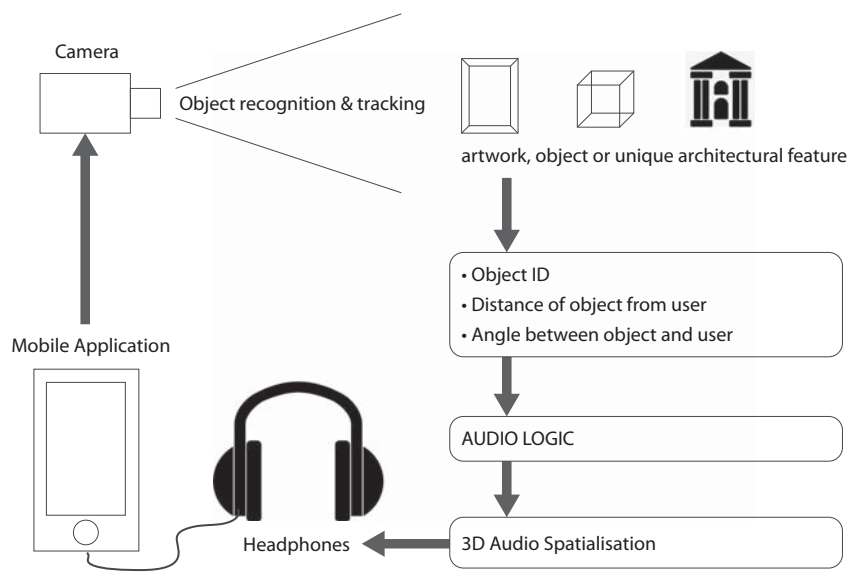

Fig. 1 System architecture of prototype

Additionally, the system is capable of determining the listener's current focus by returning the angle and distance of the listener in relation to the tracked object. An additional and important feature of this camera-based IPS is made possible through Vuforia's Extended Tracking or Simultaneous Localisation and Mapping (SLAM) capability, delivered through either Apple's ARKit [1] or Google's ARCore [15], when compiled for delivery as either an iOS or Android application respectively. Vuforia's extended tracking enables the continued recognition and estimated location of a tracked object outside of the camera's field of view. This fusion-based sensing technology extends our ability to determine the location of our physical objects and their associated virtual audio sources in relation to the listener's position in space. By being able to estimate both the angle and distance of the virtual audio sources around the listener, we can deliver a virtual and interactive three-dimensional soundscape based on the listener's physical, real-world environment.

Initial prototype designs centred around tracking the objects to which the virtual sound sources were going to be attached to, and using these as reference points to determine our listener's position and orientation, an approach that seemed natural given that these were the objects that we wanted to detect. But through the prototype development stages, once a system had been developed that demonstrated a useable degree of accuracy and reliability, and through the trials and manipulations involved in sculpting the positions and dimensions of the virtual audio sources in physical space, a 'natural feature' detection approach emerged. This approach involved providing the object tracking software (Vuforia) with isolated images of unique and static physical features within the experience environment, and determining the listener's position and orientation in relation to these physical features, and in turn determines the position of the user in relation to the object to be augmented with sound.

## 6 Authoring and development

Naphtali and Rodkin [24] define a core set of components required to construct an AAR system. These include: sensors, control methods, rules and conditions and a delivery mechanism. Within this particular AAR system, we can define our sensor component as being a camera, which will provide real-time tracking of our listener's position and for recognising environmental elements. Our control methods are virtual colliders, authored zones of space in the virtual environment, the position of which in the real world physical environment can be determined by our sensor component. These colliders act as triggers for our rules and conditions, which are essentially the authored logic that determines the audio content delivery. The delivery mechanism, the device with which our listener will interface with system, comprises of a smartphone and headphones, the former capable of realising our core set of system components either via an installed application or intrinsically via its hardware and software, the latter capable of delivering personalised, high-fidelity, three-dimensional sound.

An image target, in the form of a QR code, was uploaded to Vuforia where the image feature points are extracted and stored in a database. This image target was included as a game object within the Unity scene, with another game object added as a child of this image target object, to represent the virtual audio source. This child object was positioned virtually in relation to its parent image target to reflect the actual required position of the virtual sound source in our real-world environment (see Fig. 2).

The authored FMOD audio event was attached to this child game object, along with a collider object for triggering it. Key to this authoring approach working in relation to the designed model of spatial interaction (Fig. 3) is the use of collider components on both the virtual audio event triggers and on Vuforia's ARCamera object. The addition of a rigid body component on the latter, combined with these collider components, renders our user's mobile camera position within both the virtual and physical world of our AAR application much the same as a first-person perspective player within a video game, and, as such, other similar game-orientated authoring approaches can be adopted within FMOD. The approach of commandeering game authoring techniques, specifically collision detection, for spatial augmented experiences, is utilised and reflected upon by Greenhalgh and Benford [16] in their model of spatial interaction for a remote teleconferencing application.

## 7 Spatial interaction

The appropriation of the VR authoring technique of collision detection through the placing of collider components around the virtual sound sources and the ARCamera object begin to realise a model of spatial interaction with similarities to Greenhalgh and Benford's [16] *Aura*s, spatial zones around objects that define their region of interaction with other objects. Similarly, this approach enables an awareness of these objects to each other, indicated by their position and orientation. This awareness can be used to design a model and author a subsequent experience that can take advantage of this information to determine a user's current focus within the system, and to allow an object to determine if it is the current point of focus.

The design of focal length and width for individual virtual sound sources within the model can be achieved through the dimensions of both its range and its associated collider, the shape of its directivity pattern and through the attenuation of its signal based on the parameters of distance and angle between it and a listener. Again, this echoes [16] and the concept of the nimbus feature of an object as both a focal and advertising determiner.

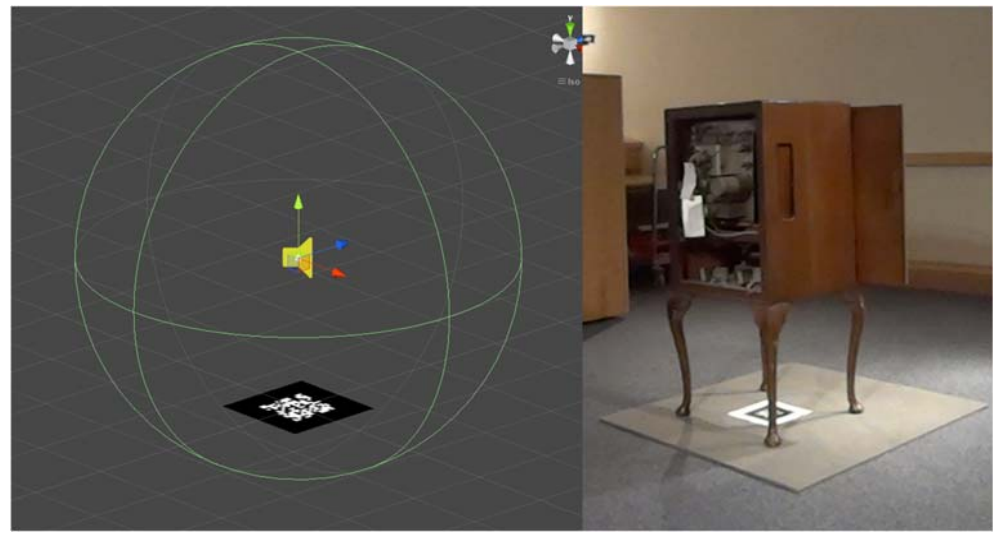

**Fig. 2** On the left, we see the virtual environment during development, showing the position of the virtual audio source and its collider component in relation to the position of the tracked image. On the right, the position of the tracked image in relation to our radio object in our real-world installation environment. The speakers of the radio were situated in the bottom of the main body of the radio unit

By adjusting the audible presence of virtual sound sources based on listener's proximity and orientation, we can design an element of focus into the experience where individual sound sources and objects can be identified and coherent and curatorially useful soundscapes can be composed. Additionally, this audible presence could be manipulated, or focal range extended, in order to give specific sources priority, or to enable them to advertise their presence more vocally than other sources within the experience.

It is these points that are of particular interest as they constitute a manipulation of the usual, or expected, attributes of a physical sound source. According to the normal physics of sound, these sound sources would continue to emanate through the soundscape, with only the altered characteristics virtually attributed to them by the game engine's audio spatialisation effects, perhaps through means of occlusion, change of environment, volume and position within the game or experience. We may see here the emergence of a model for spatial audio interaction for use within applied AAR systems, where a considered compromise is brokered between audio reality and a functional and coherent application through spatial interaction.

In Fig. 3, we see the spatial audio interaction design for the Listening Session study. In the centre is the physical vintage radio artefact, which is represented in the virtual space by the QR code on the floor below it (as shown in Fig. 2). There are four looped virtual archive radio broadcasts positioned around the QR code image target; these are positioned at 0°, 90°, 180°, and −90° and are indicated by areas A, B, C and D on the diagram respectively.

For the purposes of this initial study, a 1950s television and radio receiver were selected from the museum's collection, and contemporaneous archival radio broadcast material was obtained from an online Internet archive resource. This material included a science-fiction radio drama, a live concert hall musical performance recording, the narrated introduction to a religious music programme and an episode from a detective drama serial. All the chosen audio content was historically and geographically accurate in relation to the chosen radio receiver from the museum's collection. In addition to the recorded archival radio broadcast audio content, various recordings of radio static were obtained by recording the output from an out-of-tune contemporary radio receiver. Table 1 shows the included audio content and details of their attributes.

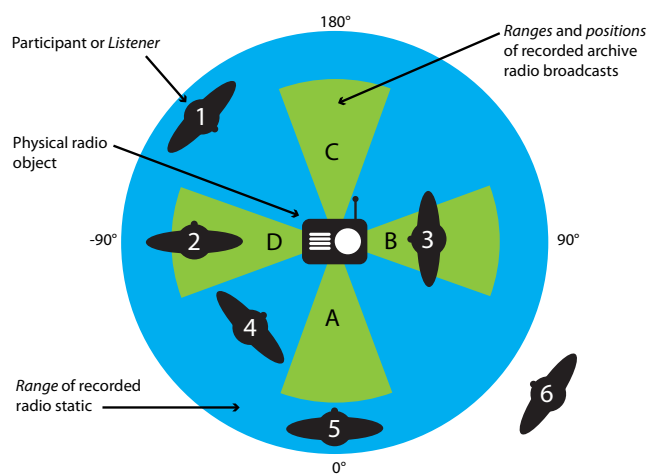

**Fig. 3** The spatial audio interaction design for the Listening Session study

The real-world positions of these virtual archive radio broadcast transmissions are achieved within the FMOD event authoring environment by cross-fading from the background radio static sound to the appropriate archive recording when the listener is in the relevant position in relation to the tracked

Table 1 Details of the audio content included in the installation

| File name | Description | Type | Function | Position | Range | Loop length |
| --- | --- | --- | --- | --- | --- | --- |
| Paul-temple.aif | Crime drama | Spoken word | Archive content | 0° | 2 m | 02.05 |
| Chapel-in-the-valley.aif | Religious music programme | Spoken word and music | Archive content | 90° | 2 m | 00.42 |
| Variety-bandbox.aif | Live recorded musical concert | Music | Archive content | 180° | 2 m | 01.56 |
| Red-planet.aif | Science-fiction drama | Spoken word | Archive content | −90° | 2 m | 02.17 |
| Static.aif | Untuned radio static | Sound effect | Transitional ambience | n/a | 3 m | 00.06 |

QR code. The cross-fading between these two audio sources is extended by 10° in each direction from its centre position, with a further 10° transitionary non-linear cross-fade to allow for a degree of comfortable, and smooth transitional listening, so small body movements do not result in sudden loses of the perceived broadcast signal. The audio sources were positioned around the radio in this fashion to promote 360° exploration of the physical artefact, and so that multiple audio sources could be tuned in to through the embodied interactions of the listener around one augmented source. The fine tuning of the cross-fade angles were a result of trial and error in the authoring process in an attempt to create smooth and seamless auditory experience, and to try and emulate the tuning of an analogue radio dial with bodily movement.

It is the listener's focus, along with their position in relation to the virtual sound source, which is situated in the same physical location as audio augmented object, that additionally determines the delivery of the audio content to the user. Within the context of this study, and the associated interaction model, the listener's focus is determined by the angle of their handheld iPhone in relation to the tracked image target. It is this, in addition to their bodily position in relation to the tracked image target, that provides a spatial interactional model that encompasses degrees of listener position, proximity and focus.

These three spatial interactional variables (position, proximity and focus) and their associated outcomes in terms of audio content delivery for the respective listener can be illustrated through a closer inspection, and a comparison of the positions of listener 2 and listener 3 in the interaction design diagram (Fig. 3).

In Fig. 3, we see listener 2 at a position of −90° in relation to the radio object and the tracked target, and therefore currently at a position where they can hear broadcast D, in contrast to listener 3, who is at a position of 90° in relation to the radio and therefore can currently hear to broadcast B.

It is the listener's proximity to the radio object that also determines if they are currently within hearing range of the broadcasts located at their current positions, and the degree to which the broadcast's signal is attenuated and mixed with background static. We can see that both listener 2 and listener 3 are within range of broadcasts D and B respectively, and therefore are able to hear these broadcasts, though listener 3's closer proximity to the object means that its signal will be less attenuated than listener 2, who is further away.

Our last interactional variable that of focus is illustrated by both listener 2 and listener 3 within Fig. 3. The focus variable is determined by the angle of the listener's device in space (in this case their handheld iPhone) in relation to the position of the tracked object. We can see that listener 2 is facing away from the radio, with it situated on their immediate right-hand side, and as a result will perceive the spatialised audio content as being emitted from their right-hand side (the direction of the radio). In contrast, we see listener 3 directly facing the radio, who, as a result, will perceive the virtual audio sources as emanating from directly in front of them.

In light of this explanation of the spatial interactional variables of position, proximity and focus, we can determine the differences in the delivery of audio content for all our listeners' locations in Fig. 3. Perhaps notable here is listener 6, who, although directly facing the radio, will hear nothing as they are well outside the range of both static and broadcast. Similarly, we see that the location of listener 5 determines that, although they are within range of the static, with the radio directly in from of them, they are beyond the range of the broadcast.

Furthermore, the delivery of audio associated to the radio object to listener 2 in Fig. 3 has the potential to encourage engagement with the object by tempting their focus, but additionally leaves them open to impressions of other potential virtual sound sources within the context of an experience with multiple audio augmented objects.

Both the real-world distance and angle between these virtual sound sources and the user can be accessed as parameters within FMOD in order to author adaptive transitions in the delivery of the audio content. This is achieved in the same way a player's character may experience virtual sound sources when exploring the virtual domain of a video game, or the way in which instrumentation within an adaptive soundtrack may be manipulated in relation to the player's health, or the proximity of enemy characters.

## 8 The study

The Gallery Listening Sessions were a set of workshops exploring the question of what 'sonic engagement' should mean, and how it should be achieved in the context of museums of science and technology. Interested parties were invited to take a guided tour of the museum's collection stores and take part in a small number of workshops. After the museum tour, attendees were invited to take part in our AAR study.

A total of 10 attendees to NSMM's Gallery Listening Session participated in the AAR study, and these participants were reflective of the Gallery Listening Session attendees in general, researchers, museum professionals, museum visitors and members of the public, of mixed age and gender. Participants were handed the iPhone and instructed to wear the headphones, ensuring they were on the correct way around, and to explore the radio object, no additional information regarding what would happen, how the technology worked or what they could expect was provided.

Due to the developmental nature of the application, participants were provided with iPhones with the mobile application already installed, and the appropriate application was either started before handing over the iPhone to the participant or pointed out to the participant amongst the other app icons on the iPhone's home screen. So that the social interactions between users could be observed and recorded participants were instructed to experience the installation in pairs. This was facilitated by having two iPhones with headphones attached available for them to use. Participants either self-organised themselves into pairs or the pairings were the result of their availability to participate either having completed other workshop activities or having completed the required ethics paperwork and consent documentation.

Both video and audio recordings were captured of the participants just prior to, during and after their engagement with the study. Participants were given the opportunity to provide both verbal and written feedback relating to their experience of the study subsequent to their participation. Verbal feedback was captured on the video camera and took the form of an open-ended discussion. Written feedback was collected on feedback forms; these were completed anonymously by participants as free text in an effort to encourage the collection of honest thoughts and descriptions from participants relating to their experience that they may have felt less willing to disclose during discussion.

## 9 Findings

Participants' written feedback was prompted by the question How would you describe your experience with the augmented radio? Verbal feedback was captured on the video camera's microphone, with participants being asked, if they were not initially forthcoming on their own accord, what they thought about the experience they had just undertaken. The bodily interactions between all pairs of participants and the radio installation were recorded on a single, wide-angle video camera that covered the interactional setting of the installation. From this view, participants were recorded entering, interacting with and leaving the setting of the installation. These recorded interactions were then indexed and thematically analysed.

In the written feedback, all but one of our ten participants described their experience as being either 'interesting' or 'fascinating'. Two participants commented on the authentic 'valve warm sound' and the 'period appropriate programming', one commenting that 'It was interesting to have new technology used to interpret a story about an older object' and that they would like to see this technology used throughout museum.

Two participants made direct references to how their bodily movements were tuning the radio into the different broadcasts, and likening this to their practical experiences and memories of tuning a traditional radio receiver. There were comments made about being able to listen to individual broadcast material, as well as being able to construct or compose an individual soundscape experience from the different elements available, 'picking up and losing the sounds'.

Additional positive references were made to the exploratory nature of the experience and its potential for being adapted as a maze, puzzle or mystery solving experience. One participant mentioned that they would have liked additional visual or textual information displayed on the phone's screen to complement and provide information about the audio they were currently listening to. Furthermore, this feature was suggested as an additional means of navigation within the experience, to visually indicate the whereabouts of specific sounds or, if you miss something, provide a means by which it could be easily found again.

In relation to the verbal feedback, participants identified with the experience of using their proximity and their position in relation to the radio to find the broadcast material amongst the sound of static as being a metaphor for what it may have been like, or what it was like, to originally tune this type of analogue radio receiver. One participant commented:

> It reminded me of how difficult and frustrating it used to be to tune a radio, because walking around the object was like tuning it.

Mentioned again in relation to the evoking of memory was the 'Faithful reproduction of the warm valve sound' indicating the potential importance of historical accuracy in the sonic delivery of the audio augmented object. Participants also expressed an interest in further levels of sonic engagement with the object, for example one participant mentioned that they almost expected to hear 'more stations when pointing the phone at the tuning dial on the radio'. Two participants made reference to the 'abstract' nature of the experience and expressed interest in having a more literal and faithful relationship between the object and the delivery of the audio content. One participant commented on how the combination of the real object and the virtual audio triggered their imagination, much like listening to music being a catalyst for the mind's eye, but suggesting that having a physical object in front of them which directly related to the content on their headphones in some way amplified this experience:

> It just brings the sound out more, so you're kind of just looking at the object, imagining things, the object's actual sounds but without touching it.

A thematic analysis of the recorded video footage of all our participants' interactions with the installation highlights a common interactional sequence as illustrated in Fig. 4. Generally, we observe eight distinct phases of interaction with the experience: preparation, familiarisation, exploration, investigation, focussed listening, second-level focussed listening, interruption and finishing. We see how, through a process of familiarisation, our participants quickly associate their bodily movements to the receipt of the spatialised audio sources, and then begin to explore the interactional setting to see what they can find. Subsequent to this, we witness our participants returning to investigate the location of some of these sources and engage in listening to them. This phase of focussed listening can sometimes result in a more attentive and engaged listening activity, observable by participants attempting to achieve a very close proximity to the location of the virtual sound source. We see how personal space and acceptable social proximities affect the process of virtual sound exploration and investigation, and how these predefined and mutually agreed proximities become more flexible during phases of engaged listening. We will now look at each identified interactional phase in a bit more detail.

## 9.1 Preparation

It is envisaged that the application will eventually be made available for listeners to download onto their own devices, enabling institutions to economically deploy such experiences. As such, familiarity and access to an appropriate device would be assumed, with the exception of the listening station approach discussed earlier. Although all participants automatically put on their headphones when they were ready to start, four participants needed to be reminded to put their headphones on the correct way around (essential for the correct orientation of the interactive surround sound). Observable from the recorded video of our participants' interactions with the installation, we notice that two out of our ten participants required instructional prompts from the researcher to engage in an exploration of the space.

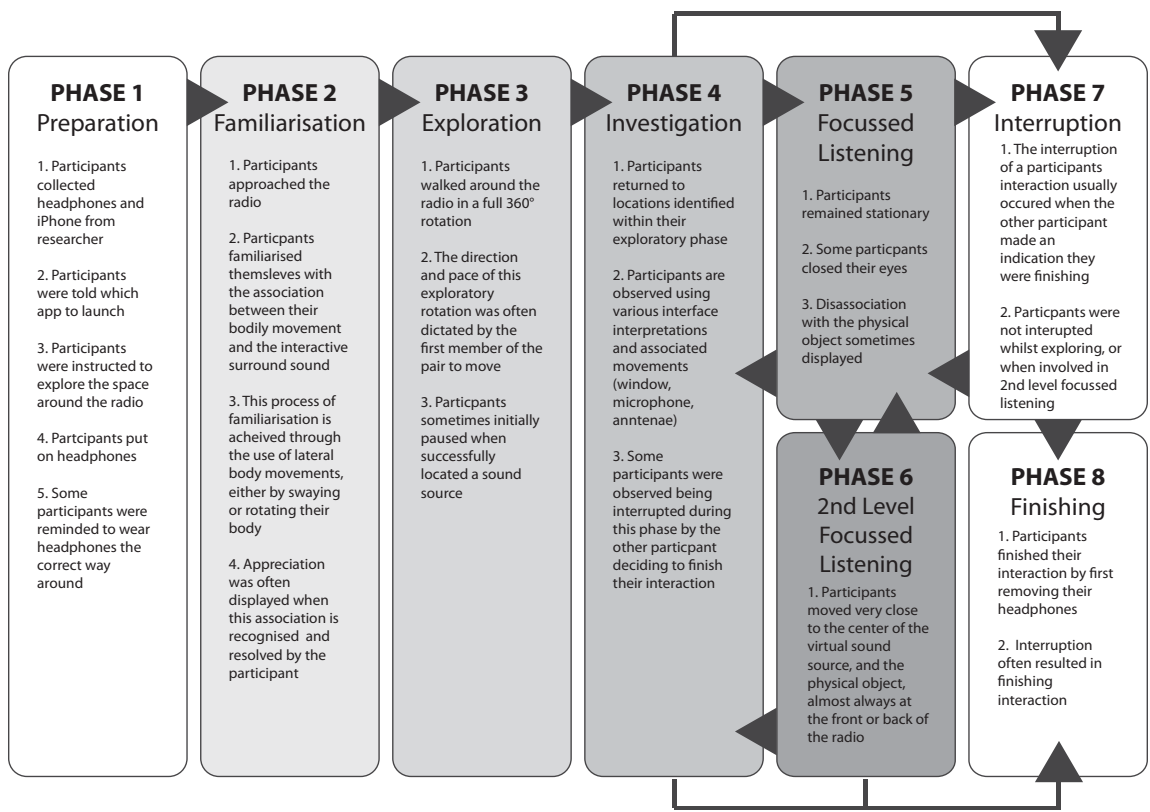

**Fig. 4** The identified different phases of interaction within the Listening Session study and their relationships to each other

## 9.2 Familiarisation

This phase of familiarisation is distinguishable within the video recordings of our participants' interactions by the various lateral movements our participants made. This seems to indicate an initial process of familiarisation with the association between bodily movement and the interactive positioning of the surround sound. These movements are often terminated by an acknowledging sign of appreciation, perhaps confirmation that the association has been recognised and understood. These lateral movements were observed being performed in a variety of different ways. Some participants swayed from side-to-side with their device held in alignment with their body and head. One participant waved their device in a lateral motion within a few moments of starting the experience and kept their body stationary whilst doing so. Another participant rotated their upper body in a lateral motion, and therefore also the device they were holding.

During this phase of familiarisation, a detachment of the focal gaze from the screen of the device was observed. In other words, the participant, through their particular process of positional familiarisation, was observing the physical object directly, rather than secondarily through the screen of the device.

This initial process of familiarisation of embodied interactions with spatialised audio via repeated lateral movement is consistent with Heller and Borcher's AudioTorch [18]. Equally consistent with AudioTorch is the way in which it is capable of achieving a quick link between the hand and ear, a link that, in most part, remains unbroken and which can be observed by participants keeping their head aligned with the orientation of the device in their hand for the duration of the experience.

## 9.3 Exploration

After the brief familiarisation phase described above, all our participants can be observed within the video recordings of their interactions walking around the radio a full 360°, often pausing briefly at the locations of the audio signals, as indicated in Fig. 3. The direction of exploration, clockwise or counter-clockwise, most often determined by the first participant to start moving around the object, equally the length of the participants' pauses at the locations of the audio signals were often determined by one participant resuming their exploration around the radio and prompting the other to resume theirs. This behaviour leads to each member of our pair of participants exploring adjacent locations of the sound source, as one member begins to travel to the location of the next broadcast, so does the other member.

This type of exploratory behaviour is observed amongst all our participant pairs, though there are some occasional exceptions. These exceptions appear to take place either when one of the participants has become engaged in the next phase of investigatory interaction, or if the participants appear to have a greater degree of social familiarity with each other, which can be indicated by an observed acknowledgment of each other and a sharing of an appreciation of the experience.

## 9.4 Investigation

Within this phase, we saw members returning to the locations of the audio broadcasts that they identified during their exploratory phase to investigate them further. We begin to see exploratory interpretations of the smartphone device as an interface to the audio content. These interpretations take on a variety of styles, with one participant holding their device aloft in an antennae-type fashion, directly reflecting the subject of both the virtual and the physical, another uses their device as a virtual microphone, moving it towards points of interest around the artefact. Others listen through the window of the screen, or rather, observe the radio through the screen of the device whilst listening through their headphones. During this phase of interactional activity, we also observe participants sharing the same audio sources and interacting with the installation in much closer proximity to each other.

## 9.5 Focussed listening

The investigation phase, where our members revisit the virtual audio broadcasts they identified within their exploratory phase, quickly develops into focussed listening. This is discernible within our video recordings of their interactions by the participant remaining stationary for a prolonged period for the first time since beginning their interactions with the installation. Evident within this interactional phase is an apparent disassociation with the physical object itself, with participants being observed closing their eyes or seemingly focussing on other more distant objects whilst they concentrate on the audio content. This behaviour is also documented in one participant's written feedback, though it is interesting that despite the visual disassociation with the radio object, a strong sonic and physical attachment to it remains:

> It was a fascinating experience. The object came alive, I entered a new sonic dimension where I was totally immersed. (I also closed my eyes repeatedly). I was trying to understand the context of sound content, the words of the man speaking.

Again, despite this visual disassociation with the object whilst engaged in these periods of focussed listening, these events initially take place at either the front or the back of the object, areas of distinct visual interest compared with the two rather plane wooden sides, with the exposed electronic and mechanical insides at the rear, and the TV screen and radio dials at the front. This behaviour is observed despite the location of the two audio broadcasts at the sides of the object, as shown in Fig. 3.

## 9.6 Second-level focussed listening

Throughout the recordings of all our pairs of participants, we witness moments when at least one of the participants engage in listening in much closer proximity to the object, often crouching down in order to obtain a physical position very close to the centre of the virtual sound source. This happens exclusively at the front or to the rear of the object where the object's mechanical and electrical interfaces and inner workings can be seen respectively.

## 9.7 Interruption and finishing

The interruption of a participant's activities, which often resulted in them finishing their interaction, resulted from one of the pair of participants deciding they have finished. Evident throughout all the recorded interactions, in all but one of our 5 pairs of participants, the end of participation is initiated by one participant removing their headphones, which prompts the other to do the same, even though the participants never started at exactly the same time. In the one event in which this did not happen, the other participant was engaged in second-level focussed listening.

We obtained from our video recordings that on average our participants spent 3′17″ exploring the installation. The combined length of unique audio content available to listen to was 6′ 20″ (excluding the looped background static recording). Therefore, if we assume that none of our participants listened to the same piece of audio more than once, we can say that on average our participants listened to 51% of the available audio broadcast material. Only one of our participants reported a potential fault with the system.

Within this model, we see some phases of interaction that resonate with the findings and observations from some of the previously mentioned related works in this area. This includes the use of virtually attached sound as an advertiser that draws users towards the audio augmented object for closer investigation. This is identified, though not specifically exploited, by Zimmerman and Lorenz [33] and could be said to be evident within our participants' trajectories from exploration through to investigation and focussed listening. Furthermore, we see evidence of this second level of focussed listening within the work of Montan [22] with differently treated zones of reverb that are triggered upon a user's close proximity to the audio augmented object, generating a soundscape within a soundscape and the feeling amongst participants of entering into a different space from outside. Based on these commonalities, we can perhaps begin to generalise more widely across cultural applications of AAR, as well as other potential applications, and perhaps provide some foundations of a theoretical model for attraction and immersion within applied AAR experiences.

## 10 Discussion

### 10.1 Serendipity versus declarative knowledge?

Throughout their description and analysis of their deployment of the Exaudimus system, Mortensen and Vestergaard [23] iterate that their interest lies in the creation of serendipitous moments of engagement, rather than assisting the listener in the collection of declarative knowledge on the subject matter. As maintained by Truax [30] and Mortensen and Vestergaard [23], such serendipitous encounters have the ability to realise engaging cultural experiences and have the potential to extend interest in the exhibition subject matter beyond the duration of the exhibition. Such serendipitous and explorative expeditions could be likened to Debord's theory of the derive [10], a détournement where one is concerned with the potential points of departure, rather than a specific destination. Mortensen and Vestergaard [23] make reference to this type of take-away chance encounter or, recontextualisation of the familiar or seemingly mundane, that acts as a catalyst for extended engagement.

Evidenced within the quotes of our participants' verbal feedback in Section 9, as within the study conducted by Mortensen and Vestergaard [23], we perhaps see evidence of the potential role of personal memory, and the triggering of it, playing a role in realising these moments of serendipity which, in turn, result in the moments of engaged exploration demonstrated by the phases of focused listening. Additionally, we see evidence that suggests how virtual sound sources when combined with physical artefacts have an ability to stimulate the imagination and realise these moments of engaged exploration. Truax [30] explicitly attributes this phenomenon to the ability of sound to create relationships between listeners and their environment, combined with a relationship between embodied interaction and embodied cognition, the idea that bodily movement influences our process of acquiring knowledge and understanding. Furthermore, with regard to this audio-object relationship, we can perhaps look towards Schafer's [27] work on the soundscape and its composite elements, where sounds are prescribed with the ability to indicate age and reflect the state of society within which they were conceived. Such attribution gives the object the power to speak to, and engage with the visitor beyond the immediate scope of the audio content with which it has been augmented.

Truax's suggestion [30] invokes Bull's [8] observations on the use of personal portable audio systems, through which users have been augmenting their environments for decades, and perhaps point towards the importance of nomadic agency within the system, where listeners remain free to explore their own relationships between virtual sound, the physical environment and its contents. Though, evidently, we should not dismiss the ability of serendipitous experiences to increase engagement, awareness and understanding of subject matter on their own, our identified phases of focused listening, also observed by Montan [22], offer opportunities to create moments within the experience when declarative knowledge could be imparted.

Mortensen and Vestergaard suggest that within immersive exhibition environments such as these learning outcomes are not facts, rather experiences, feelings and memories. But we can, perhaps, have our cake, and eat it. By initially engaging listeners with chance serendipitous encounters, we could draw them into phases of focused listening during which declarative knowledge can be imparted. The question is how can we be sure of, or how can we maximise the chances of, the existence of these initial serendipitous encounters?

## 10.2 The functional and contextual aura

We discussed previously the concept of the aura in terms of its functional role within the model of spatial interaction, namely its role in determining how individual sound sources within the soundscape communicate their presence to the listener at a systematic level [16]. But we can think of an aura, perhaps in the more traditional sense of something having an aura, as the perceived meaning of a specific object or location. For MacIntyre et al. [21], the aura of an object or location is a combination of its cultural and personal significance. But in order for a beholder to understand an object's cultural significance, they need to have somehow acquired that knowledge about the object or place; a field is perhaps just a field, until you know that it is, in fact, a battlefield.

With this view, giving an object the ability to communicate information about itself to the listener gives it the ability to extend its aura, or perhaps its ability to have a perceived aura, by communicating its cultural significance. This is perhaps of interest and importance when we start thinking about how to capitalise on serendipitous encounters within the system, and imparting declarative knowledge through them. Within the structure of a dual or multi-focal model, we could think of serendipitous encounters with an object's aura as being those of personal significance, and the subsequent encounter being that of obtaining declarative knowledge of an object's cultural significance.

Participants expressed an interest in further levels of sonic engagement with the object, which point towards a possible macro and micro focus approach within the design of the spatial audio interactional model. For example, one participant mentioned that they almost expected to hear 'more stations when pointing the phone at the tuning dial on the radio'. These reports seem reflective of the findings of Montan [22] in relation to the design of different 'acoustical zones' within the context of a single AAR subject for increasing immersion and engagement, where there was a reported impression of entering into the subject, when moving from one zone to another. Such an approach is consistent with the work of the artist Vicky Browne, where the elements of Browne's sound installation Cosmic Noise are described by Kelly [20] as having 'micro-ecologies', where the work can be listened to as a whole, or attention can be focused on certain elements to reveal 'specific and often minute sounds'.

The use of embodied interaction as a metaphor for tuning into the radio, along with historical audio realism, may provide another approach to answering the question of how to evoke personal and emotional relationships with objects. Two participants mentioned how it reminded them of their direct and personal experience of tuning in an analogue radio receiver. Additionally, the 'Faithful reproduction of the warm valve sound', may have helped to attach the virtual to the physical and constitute an increase in engagement with the artefact that is a direct result of the audio augmented reality experience.

## 10.3 Artefact as interface

Furthermore, and more specifically related to the methods and rationale involved within the design decisions of Mortensen and Vestergaard's practice-based approach [23], we see how an experimental study approach is deployed as a means of exploring the potential ways in which archival sound content could be accessed in an accessible and engaging manner. This is advocated for through the analogy of 'an informational amusement park of the future', within which there is an emphasis on maximizing the possibility of the occurrences of serendipity (defined as unexpected discoveries) as a means of promoting awareness, engagement, reflection and inspiration through the experience and exploration of embodied interaction with sound, rather than the explicit gathering of declarative knowledge.

As with the radio-based installations presented here, this is achieved by using the body like a tuning dial on an analogue radio set, allowing the visitor to find clear signals of archival content amongst the sound of static. Bijsterveld [4] describes this as a 'highly original framing of the exhibition sounds' and one where 'the exhibition space itself mimicked the technology behind the sounds that were the topic of the exhibition'. One could also argue that this act of embodied, interactive tuning constitutes a physical contextualisation of the virtual digital archive content.

This physical contextualisation is extended through the construction of listening situations, where the settings of the original physical listening environment associated with the different pieces of audio content (an armchair for content programmed in the evening, a car seat for drivetime content and a bedroom for teenage content) are reconstructed within the gallery space. Again, we see an exploration into how the material can be used to promote and focus engagement with the immaterial, a mixed reality exercise in the contextualisation of the virtual with the physical.

This approach is largely justified by an understanding that learning associated with immersion is experience driven [23]. As such, the authors anticipated visitor learning outcomes to include experiencing situations, feelings and memories, not hard facts. This approach seemingly bears fruit in the form of positive participant feedback in relation to awareness, interest, engagement and the evocation of memories associated with the audio archive content used within the exhibition. The authors admit that there is no evidence that the experience would inspire further engagement with the archive beyond the scope of the exhibition, and that there were significant problems in getting visitors to physically interact with the assembled listening situations, for example actually sitting down in the armchair. It was the embodied and intimate interactions with these assembled physical situations within the gallery space that were required in order to effectively trigger the playback of the associated archival audio content, and as such the issue of exhibition competence in relation to the hands-on engagement that this type of approach relied upon remains.

### 10.4 Hands-free, heads-up

It is the commandeering of the smartphone as a delivery mechanism that, to a great degree, enables the potential permeable nature of the experience we discussed in the previous section. One can easily imagine the impractical nature of loaning multiple VR headsets, or other less ubiquitous or expensive pieces of equipment to visitors as they wander around either inside or outside the confines of the gallery or museum space. It is the smartphone that facilitates an accessible experience, an accessible experience which is facilitated by both ease of deployment from the point of view of the institution, and ease of use from the point of view of the user.

The current prototype AAR mobile application deployed in the studies presented here can be successfully installed on most Android and iOS smartphones of up to 6 years in age, with a set of stereo headphones being the only additional piece of equipment required. Although we should not assume that every visitor would be carrying a compatible smartphone, the prevalent ownership of this enabling technology [25], along with the lack of reliance on the type of background infrastructures evident in some of the previous gallery-based AAR experiences that we discussed earlier [2, 29, 33], affords a greater amount of inclusivity and accessibility for both the visiting public and the institution within which it is deployed.

The need for only relatively dated technology, in smartphone terms, permits the somewhat inexpensive deployment at an institutional level should an even greater level of accessibility want to be provided through the use of listening stations, similar to those in the Damm Project [4]. It is envisaged that, although forgoing the true nomadic nature of the unfettered experience that has been described, the installation of listening stations in a gallery space could provide 360-degree scenes of the soundscape from a stationary position from which individual audible components of the virtual soundscape could be discerned along with their physical counterparts.

By primarily concerning ourselves with an audible experience over a visual one, we not only place less demand on, and the need for, more technologically advanced and potentially expensive resources but also make such an experience accessible by more people, and more deployable, in economic terms, for the institution. Such deployments of AAR experiences within cultural institutions have the potential to bypass situations such as a queue of visitors waiting to have a go on a limited number of VR headsets. This example is included amongst other experiences that render themselves prohibitively exclusive through the use of technology that is not ubiquitous in the public domain. In short, were visitor's own technology is not capable of, or considered within, the deployment of interactive and immersive experiences within cultural institutions.

## 11 Conclusions

By way of a conclusion, we see evidence of how contextualised and embodied interaction, along with authentic audio reproduction, can evoke personal memories associated with a museum artefact, and we see participants express interest in the acquisition of declarative knowledge, based on these initial engagements with the subject matter. Additionally, there appear practical ways in which this can be achieved through a dual, or multi-layered, focal structure, through the adoption of an aura-based functional and theoretical model, and our observations suggest that users would engage with such an approach.

Overall, we demonstrate the potential of the physical object's role as an interface for engaging users with associated virtual audio content. Furthermore, we demonstrate an initial prototype system that has the potential to impart declarative knowledge to users by exploiting initial serendipitous encounters, and present ways in which this specific capability could be extended and refined. Additionally, we observe how users become less aware of each other's presence as they become more engaged with the audio content and, as a result, initial social constraints become more flexible as participants become more engaged. Finally, by assigning real-world object-specific virtual spatialised audio sources, we demonstrate how these objects could communicate and engage beyond their traditional confines of line-of-sight within the context of a larger collection of objects, and how visitors can be drawn to engage further, beyond the realm of their original encounter.

## 12 Further work

Though, arguably, this approach performed well in teasing out initial findings to inform prototype development, and as a means through which engagement could be initialised, it stands mainly as a catalyst through which a secondary level of exploratory engagement could be initiated. Some other ideas for further work based on these findings could include ways in which the preparational and familiarisation phases of interaction could be combined through interactive spatialised audio instructions, potentially shortening the user's route to engagement. This could be considered along with how a user's personal preferences and data could combine with audio meta data in order to promote moments of personal attachment and memory by generating personalised auras within the experience, thus helping the dissemination of declarative knowledge.

**Acknowledgements** The author would like to thank Annie Jamieson and the rest of the staff at the National Science and Media Museum, Bradford, UK.

**Funding information** The author is supported by the Horizon Centre for Doctoral Training at the University of Nottingham (RCUK Grant No. EP/L015463/1).

## Compliance with ethical standards

**Conflict of interest** The authors declare that they have no conflict of interest.

**Publisher's note** Springer Nature remains neutral with regard to jurisdictional claims in published maps and institutional affiliations.